\documentclass[acmlarge]{acmart}%\documentclass[sigconf,authordraft]{acmart}
\makeatletter
\newcommand{\confshort}{\acmConference@shortname}
\newcommand{\conffull}{\acmConference@name}
\newcommand{\confdate}{\acmConference@date}
\newcommand{\confloc}{\acmConference@venue}
\AtBeginDocument{
  \fancypagestyle{firstpagestyle}{
    \fancyhead{}%
    \fancyfoot[C]{}%
  }
  \fancyhf{}
  \fancyhead[LO]{\@headfootfont\shorttitle}%
  \fancyhead[RE]{\@headfootfont\@shortauthors}%
  \fancyhead[LE]{\@headfootfont\footnotesize \confshort, \confdate, \confloc}%
  \fancyhead[RO]{\@headfootfont\footnotesize \confshort, \confdate, \confloc}%
  \fancyfoot[C]{}%
}
\makeatother
\acmBooktitle{\conffull\@ (\confshort), \confdate, \confloc}
\AtBeginDocument{%
  }

\copyrightyear{2026}
\acmYear{2026}
\setcopyright{cc}
\setcctype{by}
\acmConference[FAccT '26]{The 2026 ACM Conference on Fairness, Accountability, and Transparency}{June 25--28, 2026}{Montreal, QC, Canada}
\acmBooktitle{The 2026 ACM Conference on Fairness, Accountability, and Transparency (FAccT '26), June 25--28, 2026, Montreal, QC, Canada}
\acmDOI{10.1145/3805689.3812420}
\acmISBN{979-8-4007-2596-8/2026/06}
\begin{document}

\title[Relative principals, pluralistic alignment, \& the structural value alignment problem]{Relative Principals, Pluralistic Alignment, and the Structural Value Alignment Problem}

\author{Travis LaCroix}
\orcid{0000-0002-1724-3434}
\affiliation{%
  \institution{Durham University}
  \city{Durham}
  \state{England}
  \country{United Kingdom}
}
\email{travis.lacroix@durham.ac.uk}

\renewcommand{\shortauthors}{LaCroix}

\begin{abstract}
  The value alignment problem for artificial intelligence (AI) is often framed as a purely technical or normative challenge, sometimes focused on hypothetical future systems. I argue that the problem is better understood as a structural question about governance: not whether an AI system is aligned in the abstract, but whether it is aligned enough, for whom, and at what cost. Drawing on the principal-agent framework from economics, this paper reconceptualises misalignment as arising along three interacting axes: objectives, information, and principals. The three-axis framework provides a systematic way of diagnosing why misalignment arises in real-world systems and clarifies that alignment cannot be treated as a single technical property of models but an outcome shaped by how objectives are specified, how information is distributed, and whose interests count in practice. The core contribution of this paper is to show that the three-axis decomposition implies that alignment is fundamentally a problem of governance rather than engineering alone. From this perspective, alignment is inherently pluralistic and context-dependent, and resolving misalignment involves trade-offs among competing values. Because misalignment can occur along each axis---and affect stakeholders differently---the structural description shows that alignment cannot be ``solved'' through technical design alone, but must be managed through ongoing institutional processes that determine how objectives are set, how systems are evaluated, and how affected communities can contest or reshape those decisions.
\end{abstract}

\ccsdesc[500]{Computing methodologies~Philosophical/theoretical foundations of artificial intelligence}
\ccsdesc[500]{Computing methodologies~Machine learning}
\ccsdesc[300]{Applied computing~Sociology}
\ccsdesc[500]{Applied computing~Economics}
\ccsdesc[500]{Social and professional topics}
\ccsdesc[300]{Computing methodologies~Distributed artificial intelligence}
\ccsdesc[300]{Applied computing~Psychology}
\keywords{the value alignment problem, structural value alignment problem, principal-agent framework, axes of value alignment, misaligned objectives, asymmetric information, relative principles, shareholders, stakeholders, multi-agent coordination problems, power dynamics, scaling hypothesis for value-aligned AI, governance}

\received{13 January 2026}
\received[revised]{25 March 2026}
\received[accepted]{16 April 2026}

\maketitle

\section{Introduction}

    The value alignment problem for artificial intelligence (sometimes called ``AI alignment'' or ``agent alignment'') is commonly described as the challenge of ensuring that AI systems act in accordance with human intentions or values \citep{Russell-2019, Christian-2020, Gabriel-2020}. However, this standard description underdetermines the problem since it does not specify what or whose values are (or ought to be) considered the proper {\it targets} of alignment. To ask whether a system is aligned requires understanding for whom it is aligned, to what degree, and according to what standard. This highlights that alignment is not merely a technical problem, to be solved through engineering or normative encoding, nor is it a philosophical or sociological problem of determining the ``correct'' values; instead, it is a relational and evaluative judgment that depends on whose values are prioritised, how trade-offs are made, and how degrees of alignment are measured.

    In practice, contemporary alignment research aims to address a wide range of concrete problems in existing machine learning (ML) systems via reinforcement learning from human feedback (RLHF) and robustness, interpretability, and evaluation methods for deployed models. At the same time, however, discussions of alignment have historically emerged out of theoretical concerns about highly capable, hypothetical future systems such as artificial general intelligence (AGI) or superintelligence.%
        %%%%%%%%%%
        \footnote{For example, \citet{Gordon-2023} suggests that the alignment problem ``refers to the challenge of ensuring that AI systems behave as their creators intend, especially when the systems become more intelligent and capable than their human designers \ldots [because as] AI systems become more intelligent, they may develop goals and values that differ from those of their creators, which could lead to unexpected and potentially harmful outcomes'' (p. 77). See also, \citep{Dewey-2011, Prasad-2018, Asilomar-2017, Critch-Russell-2023, Hendrycks-Mazeika-2022, Bengio-2023, Bengio-et-al-2025}. \citet{Eckersley-2019} highlights that most ``concerns in the literature about the difficulty of aligning hypothetical future AGI systems to human values are motivated by the risk of `instrumental convergence' of those systems'' (p. 10). See, e.g., \citep{Bostrom-2003, Bostrom-2014, Omohundro-2008, Yudkowsky-2011, Tegmark-2018, Hendrycks-et-al-2023, Ngo-et-al-2023, Bengio-2023}. A comprehensive survey is provided in \citet{Ji-et-al-2025}.}
        %%%%%%%%%%
    While these debates have clarified important conceptual issues, their intellectual legacy can sometimes obscure the fact that problems of misalignment already arise in present-day systems and must be addressed within existing institutional and governance structures. Similarly, discussions of alignment are typically abstract, meaning they do not provide guidance for practical solutions grounded in the real-world functioning of these systems. Although some strands of alignment research are practical and engineering-focused, the normative, governance, and institutional dimensions of alignment---e.g., how competing stakeholders define alignment objectives, how evaluation standards are set, and how accountability mechanisms operate---remain comparatively under-theorised.

    Recent scholarship has sought to refine the standard definition by distinguishing between its normative and technical components \citep{Gabriel-2020}, differentiating inner and outer forms of alignment \citep{Hubinger-et-al-2021}, or separating forward- and backward-facing approaches to addressing misalignment \citep{Ji-et-al-2025}. While such distinctions add conceptual precision, they do not in themselves yield practical methods for mitigating misalignment. For example, identifying a 
    ``normative'' dimension of the problem acknowledges that questions of value are central; however, this acknowledgement itself offers no framework for deciding which values ought to guide AI systems or why they should be prioritised. Likewise, distinguishing between forms of misalignment is analytically useful, but it does not explain how the underlying normative issues should be addressed.
    
    This conceptual ambiguity motivates the present analysis: rather than proposing new technical alignment methods, this paper seeks to clarify the structural conditions under which instances of the alignment problem can arise across socio-technical contexts.%
        %%% Although much alignment research is highly practical and engineering-focused, the governance and institutional dimensions of alignment—such as how alignment objectives are defined, how evaluation standards are constructed, and whose interests shape these decisions—remain comparatively under-theorised. This gap motivates the present analysis. Rather than proposing new technical alignment techniques, the paper develops a conceptual framework for understanding the structural conditions under which alignment problems arise across both technical and social contexts.
        \footnote{Importantly, this perspective complements rather than replaces existing technical work. Techniques such as RLHF, interpretability tools, dataset governance practices, and benchmarking frameworks represent concrete efforts to mitigate misalignment in deployed systems. The claim here is that understanding {\it how} these techniques function within broader socio-technical environments requires conceptual frameworks that explicitly account for pluralistic values, informational asymmetries, and institutional power relations.} %
        %%% Importantly, this perspective complements rather than replaces existing technical work. Techniques such as RLHF, interpretability tools, dataset governance practices, and benchmarking frameworks represent concrete efforts to mitigate misalignment in deployed systems. Our claim is that understanding how these techniques function within broader socio-technical environments requires conceptual frameworks that explicitly account for pluralistic values, informational asymmetries, and institutional power relations.
%At the same time, many researchers who discuss alignment do so (either explicitly or implicitly) in the context of AI safety and concern themselves with speculative near- or far-future systems such as artificial general intelligence (AGI) or superintelligence. Consequently, this literature often neglects the concrete harms already produced by misaligned AI systems in the present. Finally, the imperative to {\it ensure} that AI systems align with human values encodes a form of {\it technochauvinism} \citep{Broussard-2018} insofar as this formulation assumes the inevitability (and desirability) of technological solutions to social problems.
    What is needed is clarity on value alignment as a {\it class} of problems, rather than a single, unified challenge. Instead of asking {\it which} values are the ``correct'' objects of alignment and {\it how} those values can be encoded, it is more productive to ask under {\it what conditions} value misalignment tends to arise {\it generally}. That is to say, what is the {\it structure} of the value alignment problem, and in what contexts is it truly a problem? Reframing the question in this way reveals that the challenge of aligning values is not unique to AI, but rather a special case of more general coordination problems among agents with differing goals and information.

%This paper makes three principal contributions. First, it develops a structural definition of the value alignment problem for artificial intelligence, drawing on the principal-agent framework from economics. This definition reinterprets alignment not as a singular challenge to be “solved,” but as a class of problem instances that arise whenever objectives are mis-specified or informational asymmetries distort interactions between human principals and artificial agents. 
    
%Second, it reconceptualises alignment as a socio-technical and pluralistic phenomenon, showing that once the category of “principals” expands to include multiple and heterogeneous human actors, the simplifying assumptions of classical economic models---single rational agents, stable preferences, and contractibility---collapse. The paper, therefore, motivates a shift from an economic to a socio-technical understanding of alignment, one that explicitly foregrounds power, information asymmetries, and value pluralism.
    
%Third, it extends the framework to practice, illustrating how misalignment scales with model size, data, and compute, and how it manifests in benchmark construction and evaluative procedures. By connecting these technical practices to philosophical traditions of inductive risk and epistemic humility, the paper offers conceptual and methodological guidance for mitigating misalignment in existing AI systems.

    In light of this challenge, I present a structural definition of the value alignment problem for AI, grounded in the principal-agent framework from economics. According to this model, value misalignment emerges in human-AI interactions whenever (a) the objectives of the AI system are mis-specified, or (b) informational asymmetries exist between (human) principals and (artificial) agents. Hence, misalignment can occur along three orthogonal, but interacting, axes: (1) objectives, (2) information, and (3) principals.
        %%%%%%%%%%
        %\begin{enumerate}
            %\item The {\bf objectives axis}, when proxy measures (e.g., objective functions) fail to capture the true objectives intended by human principals.
            %\item The {\bf information axis}, when model or data opacity, oversight limitations, or distributional shifts create misalignment, even if objectives are well specified.
            %\item The {\bf principals axis}, when multiple human actors, such as developers, users, and those affected by AI systems, hold divergent or conflicting values.
        %\end{enumerate}
        %%%%%%%%%%
    These three axes provide a diagnostic framework for analysing why alignment failures occur in practice. Misalignment along the objectives axis arises when proxy metrics fail to capture intended goals. Misalignment along the information axis can emerge when opacity, distributional shifts, or oversight limitations prevent principals from understanding system behaviour. Misalignment along the principals axis reflects the fact that multiple human actors---developers, users, regulators, and affected stakeholders---often hold divergent or conflicting values. Taken together, these axes reveal that alignment outcomes depend on how objectives are specified, how information flows between actors, and how conflicts among stakeholders are negotiated. This insight forms the basis of this paper's central claim: that value alignment ultimately raises questions of governance rather than engineering alone. Rather than a technical or normative problem, value alignment is primarily socio-political.

The remainder of the paper develops this argument in stages. In Section~\ref{sec:VAP}, the principal-agent framework is used to articulate the structural definition of the value alignment problem. Section~\ref{sec:Axes} further specifies the three axes of value alignment, and highlights how dynamic interactions between these axes make value alignment a fundamentally multi-dimensional socio-technical problem rather than a single technical challenge. Section~\ref{sec:Dynamics} demonstrates how, on the structural description of value alignment, increasing model generality, deployment scope, and stakeholder diversity systematically amplifies informational asymmetries, value conflicts, and power imbalances---a fact which I call the ``scaling hypothesis for value-aligned AI''. Section~\ref{sec:Governance} then draws the broader implication: because alignment failures arise from structural features of objective proxies, information asymmetries, and multi-agent value pluralism, alignment must be understood as an ongoing problem of governance. Rather than asking whether AI systems can be perfectly aligned, the more appropriate question is how institutions can manage and negotiate misalignment over time—determining, in practice, what counts as being aligned enough, for whom, and at what cost. %
    % This structural account highlights that value alignment in AI is best understood as a subset of broader multi-agent coordination problems. It also clarifies why alignment cannot be achieved once and for all: it is always relative to particular principals and operational contexts. Efforts to reduce misalignment for one group may inadvertently exacerbate it for another. Accordingly, alignment should be conceived not as a problem to be solved, but as an ongoing process of mitigation and negotiation.
    %%%
    %%%

    By situating alignment within a socio-technical framework, this reconceptualisation foregrounds the importance of factors such as power asymmetries, labour relations, and environmental impacts. Ultimately, the structural reconceptualisation reframes value alignment not as a purely technical or normative challenge but as a class of socio-dynamic problems indexed to contexts of delegation, information, and pluralism. In this sense, this paper contributes primarily at the level of conceptual infrastructure by providing a framework that helps organise and assess existing work on alignment while drawing attention to governance and institutional questions that remain insufficiently integrated into the broader alignment discourse. %This shift provides both the conceptual clarity needed to navigate alignment debates and a practical foundation for addressing misalignment in existing and future AI systems.

\section{The Value Alignment Problem for Artificial Intelligence}
    \label{sec:VAP}

    A standard instance of the value alignment problem for AI arises when a model's objective function---i.e., the mathematical specification of what the system is meant to optimise or achieve---is poorly or incompletely defined. In such cases, the ``objectives'' (``values'', ``goals'', ``incentives'', etc.) of an ML model are misaligned with the true objectives for the system---what we might colloquially call ``our values''. %This situation represents a failure of alignment between what the model {\it optimises} and what humans {\it intend}. 
    Such misalignment is not merely a theoretical concern about hypothetical systems; it already appears in present-day applications---e.g.,  when optimisation metrics fail to capture broader social goals or when deployed systems behave in ways that were not anticipated by their designers.

    In economics, law, and political science, this situation is more commonly referred to as a {\it principal-agent problem} (also known as an ``agency dilemma'' or an ``incentive problem''). In these contexts, a principal delegates authority to an agent to act on her behalf, yet the agent's actions may not always reflect the principal's best interests. This framework has become increasingly relevant in the AI literature, where scholars have drawn explicit parallels between value alignment and the principal-agent problem \citep{Hadfield-Menell-2021, Anonymous-2025}. Recent work emphasises the structural similarities between these domains, framing value alignment as a multi-agent coordination problem \citep{Fisac-et-al-2020}. These connections are reflected in a growing body of practical alignment research that uses economic and game-theoretic tools to design training procedures, feedback mechanisms, and oversight systems for real-world ML models. For example, this economic framework underpins contemporary approaches like cooperative inverse reinforcement learning (CIRL) \citep{Hadfield-Menell-et-al-2016}, the incomplete contracting framework for AI design \citep{Hadfield-Menell-Hadfield-2019}, and game-theoretic methods \citep{Hadfield-Menell-et-al-2017, Russell-2019, Garber-et-al-2025, Emmons-et-al-2025, Shah-et-al-2020} that explicitly leverage informational asymmetries to improve alignment.%
        %%%%%%%%%%
        \footnote{These approaches illustrate that alignment research today operates at the intersection of theory and practice, combining formal models with empirical experimentation on deployed systems. The contribution of the present paper is therefore not to replace these approaches but to situate them within a broader structural account of alignment that highlights how objective specification, information asymmetries, and pluralistic stakeholders jointly shape alignment outcomes in real socio-technical systems.}

\subsection{The Principal-Agent Framework from Economics}

    The principal-agent framework formalises a class of problems that can arise when one actor (the principal) delegates decision-making authority to another (the agent). The central concern is that the agent, possessing their own desires, incentives, objectives, etc., may act in ways that diverge from the principal's objectives. This problem is pervasive in human-human relations---for instance, between employers and employees, shareholders and managers, or citizens and their elected representatives.%
        %%%%%%%%%%
        \footnote{See discussion in \citet{Kerr-1975, Jensen-Meckling-1976, Eisenhardt-1989, Laffont-Martimort-2002}.}
        %%%%%%%%%%
    Hence, principal-agent models examine the problem of creating the correct incentives in a non-cooperative setting with asymmetric information, ensuring that the agent acts in the principal's best interest.

    One key feature of this framework is that informational asymmetries are fundamental to misalignment in economic contexts. Three key types of asymmetries are identified: %
        %%%%%%%%%%
        \begin{enumerate} 
            \item {\bf Non-verifiability} arises in cases where, even if actions and outcomes are observable, they may not be verifiable by third parties (e.g., courts or regulators). %
            \item {\bf Moral hazard} (hidden action) arises when the principal cannot observe whether the agent's actions faithfully serve her interests. %
            \item {\bf Adverse selection} (hidden information) arises when the principal cannot fully assess the agent's abilities or characteristics before contracting. %
        \end{enumerate}
        %%%%%%%%%%
    When information between the principal and the agent is symmetric (in the idealised economic model), the principal can theoretically design a contract that induces the agent to act in her best interest, even when their goals diverge. Conversely, even when the principal and the agent have identical incentives, informational asymmetries can cause the agent to (inadvertently) act in a way that the principal does not want. Thus, in the economic context, competing incentives (or ``misaligned values'') are {\it neither necessary nor sufficient} to generate a principal-agent problem; rather, such problems emerge from the interplay between incentive structures and information asymmetries.

\subsection{Extending the Framework to Artificial Agents}

    Viewed structurally, the principal-agent problem represents a class of problem instances that can arise either because of conflicting incentives (misaligned values) or because of asymmetric information between two (human) actors: a delegating principal and an acting agent. This framing can be productively extended to human-AI interactions to describe the value alignment problem for AI. In this context, a human agent is analogous to the principal, where a principal might be conceived of as the user, system designer, or company on whose behalf the agent acts (shareholders in the systems), or the principal may be understood as an individual affected by an AI system (stakeholders). The AI system itself is analogous to the agent.

    This insight leads to the following structural definition of the value alignment problem based on the principal-agent framework.%
        %%%%%%%%%%
        %\footnote{See further details in \citet{Anonymous-2025}.}
     %%%%%%%%%%
        \begin{quote}
            {\bf The Value Alignment Problem} (Structural Definition)\\
            A problem that arises from the dynamics of multi-agent interactions involving the delegation of tasks from one actor (a human principal) to another (an AI agent). This problem can arise whenever
                \begin{enumerate}
                    \item[($a$)] The agent's objective function is misaligned with the true objective of the principal(s); {\it or},
                    \item[($b$)] There are informational asymmetries between the principal and the agent. 
                \end{enumerate}
        \end{quote}
        %%%%%%%%%%
    Instead of focusing on the normative component of the value alignment problem (``What are the correct values to encode in AI systems?'') or the technical component (``How do we encode said values?''), this description emphasises the {\it contexts} in which misalignment can arise---i.e., the socio-dynamic structure of value alignment.

    Within this framework, the value alignment problem for AI can be understood as varying along three orthogonal but interacting axes:
        %%%%%%%%%%
        \begin{enumerate} 
            \item The {\bf objectives axis}, which describes the extent to which the system's formal objectives (i.e., objective functions) accurately capture the principal's true goals.
            \item The {\bf information axis}, which describes the degree of transparency, observability, and verifiability between the system and its human principals.
            \item The {\bf principals axis}, which describes the plurality and diversity of relevant human actors, encompassing both shareholders (developers, companies) and stakeholders (affected individuals or groups).
        \end{enumerate}
        %%%%%%%%%%
    These axes are orthogonal to the extent that a reduction of misalignment along one axis does not guarantee a reduction of misalignment along the other axes;  misalignment can arise independently along any one of them. For instance, an AI system may have a perfectly specified objective function (objectives axis) but still fail to be aligned because its outputs cannot be effectively monitored or verified (information axis). Likewise, a system may exhibit near-perfect performance and transparency yet remain misaligned because it optimises for the goals of one group of principals at the expense of others (principals axis). Each axis, therefore, isolates a distinct mechanism through which misalignment can occur.

    However, there are interaction effects between them, meaning that misalignment along one axis can exacerbate misalignment along another. For example, in predictive policing systems, the use of biased historical arrest data (a proxy problem along the objectives axis) both embeds and conceals underlying informational asymmetries between developers, data subjects, and law enforcement institutions. This feedback loop also redefines who counts as a ``principal'' in practice---since the communities most affected (stakeholders) have the least access to correcting the model's behaviour. In this sense, orthogonality describes analytical separability, not empirical independence: the three axes co-produce each other in real-world deployments. %
        %for instance, limited transparency can magnify the consequences of misspecified objectives. 
    The principals axis is particularly relevant for work in pluralistic alignment, where insights from social choice theory and deliberative democratic models are increasingly used to formalise how diverse human values might be aggregated or negotiated \citep{Conitzer-et-al-2024, Sorensen-et-al-2024}. This plurality of principals reveals that value alignment in AI is inherently multi-agent in nature. It is not merely about aligning ``AI values'' with ``human values'' in the abstract, but about reconciling diverse and sometimes conflicting human interests embedded in socio-technical systems.

    That said, a key difference between the principal-agent framework (for human-human interactions) and its analogue for human-AI interactions is that values---and therefore misalignment of values---are {\it inherent} to the human-human interaction insofar as principals and agents have inherent values. In contrast, in the human-AI interaction, there is no inherent misalignment between the (human) principal and the (artificial) agent because the agent has no inherent values: the ``values'' (i.e., objective functions) of the agent are programmed. In this latter case, competing incentives (value misalignment) can arise because it is impossible to specify an objective function completely and correctly; hence, whereas misaligned incentives are neither necessary nor sufficient for generating a principal-agent problem in economic contexts (the case of human-human interactions), they can do so in AI contexts (the case of human-AI interactions).

    The structural definition of the value alignment problem captures everything that the standard definition is intended to capture, but it provides additional clarity in thinking about how these problems arise. The issue is not {\it simply} that goals, incentives, values, etc., are misaligned, because even when objective functions perfectly represent our true goals, informational asymmetries between the principal and the agent can still give rise to an instance of the value alignment problem.

\section{The Three Axes of Value Alignment}
    \label{sec:Axes}

    Having defined value alignment structurally, we can now examine the three axes along which instances of misalignment arise: the objectives axis, the information axis, and the principals axis. Each axis represents a distinct yet interrelated source of divergence between the intentions of human principals and the behaviour of artificial agents. Understanding these axes in detail clarifies why alignment cannot be reduced to a single technical or normative question; instead, it must be treated as a multi-dimensional problem emerging from the socio-technical structure of human-AI interaction.

\paragraph{The Objectives Axis.}

    The first axis concerns misspecified objective functions, which occur when the formal optimisation target of an AI system diverges from the true objective for the system. This is analogous to conflicting incentives in the principal-agent framework and corresponds roughly to the ``outer alignment'' problem identified by \citet{Hubinger-et-al-2021}. In this sense, the objectives axis captures what most people intuitively mean by value misalignment---e.g., reward hacking, negative side effects, or perverse incentives \citep{Amodei-et-al-2016}. Thus, poorly designed objective functions can lead to instances of the value alignment problem whenever there is a conflict between the objective function and the true objective (i.e., the ``intentions'' or ``values'' specified by the standard definition of the value alignment problem).

    In a standard machine-learning model, the objective function defines the optimisation landscape over which the model learns. However, the solution space for an optimisation problem defined by an objective function includes a ``pathology'' of local optima, meaning these landscapes are frequently deceptive \citep{Goldberg-1987, Mitchell-et-al-1992, Lehman-Stanley-2008, Sipper-et-al-2018}. As a result, the objective function ``does not necessarily reward the stepping stones in the search space that ultimately lead to the objective'' \citep[p. 329]{Lehman-Stanley-2008}. Consequently, many objective functions are constructed {\it ad hoc}, privileging what is easy to measure rather than what genuinely captures human goals. %
    %%%
    %%%
    %%%
    The underlying reason that poorly specified objectives yield misalignment is that objective functions are mere proxies for the true objective. From the AI system's ``point of view'', however, the proxy {\it is} the objective---a model cannot distinguish between the proxy and the task it approximates. Moreover, proxies permeate every aspect of an ML system along the algorithmic development pipeline---e.g., at the problem specification stage, a task is a proxy for a real-world problem; during the model design stage, an objective function is a proxy for the true objective; during the testing/validation/deployment stages, training data, test data, and benchmarking datasets or metrics are proxies for real-world distributions. Because each layer of the pipeline replaces the complex, value-laden world with simplified formal representations, alignment depends critically on how well these proxies track true objectives.

    To give a concrete example of value misalignment along the objectives axis, consider again the case of predictive policing.%
        %%%%%%%%%%
        \footnote{See further discussion in \citet{ONeil-2016, Lum-Isaac-2016, Ensign-et-al-2018, Benjamin-2019, Broussard-2023}.}
        %%%%%%%%%%
    Charitably, the intended goal of such systems is to reduce crime by allocating police resources efficiently. However, the objective {\it function} operationalised in these models typically seeks to predict where crime is most likely to occur---a proxy for the (purported) true objective. Because actual future crime is unobservable and, therefore, impossible to optimise directly, the model relies on historical arrest data as a proxy. However, arrest data is a more accurate measure of {\it police activity} than {\it crime incidence}. As such, these data are heavily influenced by historical policing practices. Areas that have been over-policed in the past tend to generate more arrests, which in turn train subsequent models to allocate even more police resources to those areas. This feedback loop ensures that over-policed neighbourhoods continue to appear ``high-risk'', reinforcing systemic bias. For this reason, \citet{Benjamin-2019} calls predictive policing a ``crime production algorithm'' (83), which contradicts the purported aim of the model---i.e, to reduce crime. %
    Value alignment along the objectives axis requires that proxies be sufficiently faithful representations of true objectives, which is often difficult or impossible in the case of complex socio-cultural tasks.
        %%%%%%%%%%
        %\footnote{In some cases, such as predicting future crime, perfect proxy fidelity is impossible because the target involves inherently unknowable future contingencies.}
        %%%%%%%%%%

\paragraph{The Information Axis.}

     While misaligned objectives exacerbate these issues, they often interact with---and are compounded by---informational asymmetries between humans and AI systems, which constitute the second axis of the alignment problem. This axis corresponds to the ``inner alignment'' problem \citep{Hubinger-et-al-2021}, which encompasses issues such as scalable oversight, safe exploration, and robustness to distributional shifts \citep{Amodei-et-al-2016}. Closely related ideas have also been explored in the literature on the off-switch game \citep{Hadfield-Menell-et-al-2017}, CIRL \citep{Hadfield-Menell-et-al-2016}, and (partially observable) assistance games \citep{Shah-et-al-2020, Emmons-et-al-2025}.%
        %%%%%%%%%%
        \footnote{ \citet{Garber-et-al-2025} suggest that assistance games are (partially-observable) generalisations of CIRL, which is the underlying framework for the off-switch game.}
        %%%%%%%%%%
    In this latter framework, alignment is modelled as a cooperative game between a human and an AI system under conditions of partial observability, where the agent must infer the human's preferences from limited signals and where each party may possess different information about the environment. This literature explicitly highlights how asymmetric or incomplete information between humans and AI systems can generate alignment challenges, thereby illustrating a formal treatment of the informational asymmetries captured by the information axis described here.

    Informational asymmetries occur when the principal lacks sufficient knowledge to assess, verify, or predict the agent's behaviour. By analogy with the principal-agent problem, three primary forms of informational asymmetries show the circumstances under which an instance of the value alignment problem can arise along the information axis: %
        %%%%%%%%%%
        \begin{enumerate} 
            \item {\bf Non-verifiability} results from the principal (or an external third party) being unable to verify whether the agent's actions or outputs genuinely satisfy the intended objective. %
            \item {\bf Moral Hazard} arises when the agent has access to information unavailable to the principal, or its internal operations are unobservable, enabling behaviour that diverges from the principal's goals. %
            \item {\bf Adverse Selection} occurs when the principal has incomplete or misleading information about the agent's capabilities or internal characteristics before delegating a task. %
        \end{enumerate}
        %%%%%%%%%%
    %%%
    In the context of incomplete contracting, {\it non-verifiability} refers to situations where certain aspects of the contractual agreement are difficult or impossible to monitor or verify. In the case of the value alignment problem for AI, this type of informational asymmetry captures questions surrounding {\it scalable oversight}, underscoring the difficulty of effectively monitoring and controlling increasingly complex and widespread AI systems as they scale in size and scope.

    {\it Moral hazard} (endogenous to the principal-agent relationship) arises when the system's internal operations or learning processes cannot be fully observed. For example, {\it safe exploration} problems occur when reinforcement-learning agents must gather information through trial and error, potentially taking unsafe or undesirable actions during training or deployment. As models grow in scale and parameter complexity, their optimisation landscapes expand correspondingly, creating vast solution spaces with numerous local optima. A model might perform well on its training objective yet behave unpredictably when confronted with out-of-distribution data. In such cases, misalignment results not from a flawed objective, per se, but from the principal's inability to observe or interpret the agent's internal state. The POAG framework formalises such scenarios as ones in which the agent must act under uncertainty about the human’s latent reward function and the true state of the environment, reinforcing the role of partial observability and hidden information in alignment failures.

    {\it Adverse selection} (exogenous to the principal-agent relationship) has an AI analogue in the black-box nature of state of the art models---i.e., when the internal workings of a model are not easily interpretable or transparent to humans. Deep neural networks, for instance, encode relationships among millions or billions of parameters across multiple layers. These architectures learn high-dimensional, hierarchical representations that are largely inscrutable to human observers. Because it is difficult to trace how specific input features affect outputs, developers and users effectively contract with agents whose internal processes they do not understand. This opacity is compounded by the use of scraped or proprietary datasets, where the provenance, quality, and bias of data remain uncertain. As datasets scale, ensuring their fairness and ethical integrity becomes increasingly infeasible---again widening the information gap between human principals and machine agents.

    Hence, even when objectives are perfectly specified---i.e., there is sufficient alignment along the objectives axis---informational asymmetries can independently generate alignment failures. Highly complex systems may reach states of inherent opacity, where no feasible method of verification exists. This problem is especially acute in normative contexts, where there is no single, unambiguous ``correct'' answer to verify against \citep{LaCroix-Luccioni-2025}. Consequently, the information axis reveals that alignment failures can persist even under idealised objective specification, which is often the target of purely technical approaches to alignment.

\paragraph{The Principals Axis.}

    The third axis foregrounds the question: {\it with whose values are AI systems meant to align?} Standard formulations of the value alignment problem assume that alignment should target ``the values of humanity'', as though a unified set of human values could be identified and encoded. However, this approach is problematic on both conceptual and practical grounds.
    %%%
    %%%
    %%%
    As the number of principals whose values are under consideration increases, the intersection of their preferences rapidly diminishes. If we interpret ``human values'' as the conjunction of specific individual preferences, the resulting set is effectively empty since people value conflicting things. This suggests that to apply universally, any purported ``human values'' must be so abstract---e.g., justice, fairness, welfare---that they become too coarse-grained to serve as operational objectives for AI systems.  Thus, aligning AI systems with ``the values of humanity'' is impossible at any meaningful level of granularity. Recent work in AI alignment and human-AI interaction has begun to confront this difficulty directly by developing methods that attempt to represent plural and even conflicting human perspectives rather than collapsing them into a single aggregate signal.%
        %%%%%%%%%%
        \footnote{See, e.g., \citet{Kirk-et-al-2024, Sorensen-et-al-2024,Sorensen-et-al-Roadmap-2024, Huang-et-al-2024, Bergman-et-al-2024, Peterson-et-al-2024, Davani-et-al-2022, Gordon-et-al-2022, Tessler-et-al-2024} and discussion in \citet{Fazelpour-Fleisher-2025}.} %
        %%%%%%%%%%
    For example, jury learning approaches  explicitly model multiple evaluators whose judgments are preserved as distinct inputs rather than averaged away \citep{Gordon-et-al-2022}. Deliberative systems, such as the ``Habermas Machine'', seek to approximate structured democratic deliberation in which competing viewpoints are surfaced, debated, and revised \citet{Tessler-et-al-2024}. Related pluralistic approaches---including frameworks inspired by Overton-style pluralism, represent value landscapes as sets of permissible viewpoints rather than a single optimum aim to preserve dissent and normative diversity within AI-assisted decision processes \citep{Sorensen-et-al-Roadmap-2024, Poole-Dayan-et-al-2026}.

    From the structural perspective developed here, the principal's identity is not fixed but context-dependent. In human-AI interactions, principals can be divided broadly into two overlapping categories: %
        \begin{enumerate}
            \item The set of {\bf shareholders}, which includes those directly involved in creating, deploying, or profiting from AI systems---e.g., companies, research labs, developers, data scientists, system administrators, regulators, compliance officers, end users, consumers, etc. %
            \item The set of {\bf stakeholders}, which includes those directly or indirectly affected by these systems' creation and deployment---e.g., individuals, communities, or the broader public. %
        \end{enumerate}
        %%%%%%%%%%
    Although these groups may often overlap, the most consequential instances of misalignment will arise when stakeholders are distinct from shareholders---for instance, when an algorithm optimises for corporate profit at the expense of public welfare. Each group embodies different values, incentives, and priorities, shaping both how objectives should be specified and how success should be evaluated.
    %%%
    %%%
    %%%
    Hence, alignment must be understood as a process of negotiation among plural principals, whose values may diverge or even conflict. This recognition underscores why the value alignment problem is fundamentally socio-technical: resolving it requires not just better algorithms, but better mechanisms for adjudicating competing interests, distributing authority, and ensuring accountability in the design and deployment of AI systems.

\subsection{Dynamics and Interactions Among the Axes of Value Alignment}

    Having identified the three orthogonal axes that structure the value alignment problem---objectives, information, and principals---it is important to consider how these dimensions interact dynamically. While each axis can independently give rise to instances of misalignment, in real-world systems these axes rarely operate in isolation. Instead, they form a system of dependencies, feedback loops, and amplification effects that jointly determine the stability and direction of alignment over time. %
    %%%
    %%%
    %%%
%\paragraph{Coupling Effects.}
    %%%
    %%%
    %%%
    As mentioned, a reduction of misalignment along one axis does not necessarily entail a reduction along the others. For example, improving the specification of an objective function (reducing misalignment along the objectives axis) may require complex optimisation processes or large-scale data collection that increase informational asymmetries between the principal and the agent. Similarly, expanding the set of principals to include more stakeholders (improving pluralistic alignment) can complicate the definition of an objective function, thereby worsening misalignment along the objectives axis. These trade-offs suggest that alignment is not a static property but a dynamic equilibrium maintained within a multi-dimensional constraint space.

%\paragraph{Feedback Loops and Systemic Risk.}

    Misalignment along one axis can also generate feedback loops that exacerbate problems elsewhere. Consider again the case of predictive policing: an initial bias in data collection (information axis) leads to distorted objective functions (objectives axis), which in turn reinforce the biases affecting future data (further informational misalignment). Over time, this cycle compounds misalignment across axes, transforming what begins as a technical issue into a systemic one with significant social and ethical consequences. In this way, alignment failures can propagate through systems much like contagions propagate through networks. Addressing alignment, therefore, requires not only correcting individual failures but also understanding the causal structure and feedback dynamics among the three axes.

%\paragraph{Toward a Systems Perspective on Alignment.}

    Viewing the value alignment problem through this systems lens highlights the need for dynamic governance mechanisms that can adapt as models, data, and social contexts evolve. Static alignment solutions---those that assume a fixed objective, a stable information environment, or a uniform principal---will eventually fail as systems drift from their design conditions. Consequently, the study of alignment must extend beyond isolated technical interventions toward the design of institutions, protocols, and oversight structures capable of sustaining alignment over time. %
    %%%
    %%%
    %%%
    In light of these considerations, we turn now to a higher-order feature of the alignment problem: the plurality of human principals whose values and interests define what ``alignment'' even means.

\section{Alignment with Pluralistic Values}
    \label{sec:Dynamics}

    A fundamental challenge emerges from the structural account of value alignment: namely, that alignment cannot be conceived monolithically. Current technical approaches to alignment---most notably, RLHF---are primarily designed to align models with the average preferences of humans. While this strategy is often framed as ``democratic'', it effectively erases diversity by smoothing over legitimate differences among individual and group values \citep{Aroyo-et-al-2023, Siththaranjan-et-al-2024, Sorensen-et-al-2024}. This observation should be understood as a diagnostic claim about prevailing training and deployment incentives rather than a dismissal of emerging pluralistic alignment research. In principle, feedback aggregation methods can be designed to preserve disagreement, represent minority viewpoints, or enable deliberative synthesis among competing perspectives. In practice, however, standard deployment pipelines---especially those optimised for scalability, consistency, and product usability---tend to reward the production of a single coherent response distribution, which implicitly encourages the smoothing or aggregation of divergent human judgments unless explicit institutional commitments are made to preserve pluralism. %
    %%%
    %%%
    %%%
    Pluralistic alignment, by contrast, seeks to computationally model value pluralism itself. This approach recognises that as AI systems (especially large language models) are increasingly marketed as ``general-purpose'' technologies for use across diverse domains by heterogeneous users, the relevant value landscape becomes more fragmented and contested. The challenge is not to identify a single set of ``human values'', but to navigate among many partially overlapping and sometimes conflicting sets of values held by different principals. The structural definition brings to light how difficult this challenge is, and how contemporary machine-learning approaches exacerbate these problems.

\subsection{The Scaling Hypothesis for Value-Aligned AI}

    The structural definition of the value alignment problem clarifies that every instance of misalignment is indexed to a particular principal or set of principals. Some principals' goals (for example, those of developers, regulators, and consumers) may converge. Still, in many cases, they diverge in ways that are decisive for assessing whether a system is genuinely aligned. A model may therefore be well aligned relative to one principal while simultaneously misaligned relative to another. Alignment, in this sense, is always relative and always partial---a matter of degree rather than an absolute state. The structural definition underscores that talk of ``solving'' the value alignment problem involves a category mistake: the value alignment problem is not a problem {\it per se}, but a class of problems that can be more or less instantiated. From this perspective, the notion of alignment {\it simpliciter}---i.e., complete alignment across all axes and for all relevant principals---is incoherent for sufficiently complex systems.

    The structural account also motivates the following {\it scaling hypothesis} in the context of value alignment. 
    %%%%%%%%%%
    \begin{quote} 
        {\bf Scaling Hypothesis for Value-Aligned AI}. \\ 
        As AI systems increase in (i) model generality, (ii) deployment scope, and (iii) stakeholder diversity, the difficulty of achieving value alignment increases because these forms of scaling systematically amplify informational asymmetries between humans and AI systems and pluralistic conflicts among the principals whose values the system must serve.
    \end{quote}
    %%%%%%%%%%
    Importantly, this hypothesis should not be interpreted as a formal theorem about ML systems, but as a structural claim about how the three axes of alignment interact as AI systems expand in capability and deployment. More precisely, scaling tends to occur simultaneously along three dimensions: (i) model generality, as systems move from narrow task-specific tools toward general-purpose capabilities; (ii) deployment scope, as systems are integrated into a wider range of social and economic contexts; and (iii) stakeholder diversity, as the set of individuals and institutions affected by the system expands. Each of these dimensions enlarges the alignment problem in a distinct way. Greater model generality increases the number of contexts in which behaviour must remain aligned, often requiring deployment under conditions that differ substantially from training data or the target objectives. Broader deployment scope introduces new environments, regulatory regimes, and norms that may not have been anticipated during development. Expanding stakeholder diversity increases the likelihood that the relevant principals hold incompatible values or priorities.

These dynamics imply a structural scaling effect: as models become more general-purpose and widely deployed, both informational uncertainty and normative disagreement grow. Informational asymmetries increase because larger models trained on vast datasets and complex architectures become harder for humans to interpret, audit, or verify. At the same time, pluralistic conflict intensifies because a single system must simultaneously serve users with heterogeneous interests, cultural norms, and risk tolerances. Alignment difficulty therefore grows not merely because models become technically more capable, but because the social and epistemic environment in which they operate becomes more complex. Empirical patterns in ML development lend plausibility to this hypothesis. Large foundation models are trained on heterogeneous web-scale datasets and subsequently deployed across domains ranging from education and healthcare to journalism and governance. Each new domain introduces distinct objectives, risk profiles, and stakeholder groups. As a result, alignment failures in such systems often manifest not as simple objective misspecification but as context-sensitive conflicts among legitimate values. In contrast, highly successful systems such as domain-specific scientific models illustrate the opposite pattern: alignment is comparatively tractable when objectives are tightly specified and the set of relevant principals is limited.

    %A broader principle that is entailed by the structural definition is that {\it misalignment scales with size, data, and compute}. This ``Scaling Hypothesis for Value-Aligned AI'' follows from the fact that as models increase in size, they demand exponentially more data and computational power, which in turn amplifies informational asymmetries between principals and agents. Consequently, as we scale systems toward general-purpose functionality, we also scale the difficulty of ensuring alignment across diverse stakeholders. This implies that while small, domain-specific systems with narrow objectives (and few directly affected stakeholders) may achieve approximate alignment, larger or more general-purpose models inherently bear greater alignment burdens. 

    For example, \citet{Andrews-2023} suggests that DeepMind's AlphaFold and AlphaFold 2.0 are among the most impressive results that ML methods have achieved for science, not the least because the protein-folding problem in structural biology was largely considered intractable. From the perspective of structural alignment, part of the incomparable success of AlphaFold can be attributed to the fact that the model is not domain-generic. Namely, those systems that lend themselves to alignment {\it simpliciter} are systems with narrow targets (limiting direct stakeholders) that model readily formalisable objectives. The scaling hypothesis for value-aligned AI therefore predicts a gradient of alignment difficulty: the closer a system approaches general-purpose functionality and widespread deployment, the more the alignment problem shifts from a primarily technical challenge to a socio-technical one involving informational opacity and pluralistic value conflict.

    %Even so, as we increase the capacities of a model in alignment with one set of principals, alignment may decrease for another set of principals; this implies that mitigating misalignment along the objective or information axes can increase misalignment when the problem is indexed to a distinct set of principals. This follows from the very idea of {\it pluralism}, which is the target of pluralistic alignment. Those systems that lend themselves to alignment {\it simpliciter} are systems with narrow targets (limiting direct stakeholders) that model readily formalisable objectives.%    

\subsection{Pluralism, Power, and the Socio-Technical Context}

    Crucially, AI models do not operate in isolation. They are embedded in wider socio-technical systems that encompass diverse sets of principals---shareholders and stakeholders alike. \citet{McQuillan-2022} highlights that AI is more than a set of ML methods: It is impossible to separate the technical aspects of AI from the social contexts in which these models are created, trained, tested, and deployed. Present-day generative models, for example, have been trained on vast datasets that include copyrighted material, personal data, and cultural artefacts---often without consent or compensation. Hence, these models involve an unethical kind of labour theft \citep{Goetze-2024}. Privacy and consent violations in datasets often adversely affect individuals in marginalised communities, as attempts to ``diversify'' datasets to be more representative can incur costs to those groups concerning privacy, exploitation, monitoring, and other issues \citep{Hamidi-et-al-2018, Hoffmann-2019, Raji-et-al-2020}. These harms are compounded by the material and environmental costs of AI infrastructure---from extractive labour and resource mining \citep{Crawford-2021} to the significant carbon footprint of large-scale training \citep{Luccioni-et-al-2023, Luccioni-et-al-2022}, which disproportionately impacts the ``often lower income and thus most neglected humans in society'' \citep{Raji-Dobbe-2023}. Hence, alignment demands more than technical fixes; it requires an examination of the power relations that determine whose values count in practice.

    These socio-technical dynamics provide further motivation for the scaling hypothesis described above. As AI systems scale in capability and deployment, the set of affected stakeholders expands far beyond the immediate developers and users. Systems trained and deployed by a small number of organisations may influence millions of individuals across jurisdictions, cultures, and regulatory environments. The resulting expansion of the principal set intensifies pluralistic disagreement about acceptable uses, risks, and benefits. In other words, scaling AI systems simultaneously scales the political and ethical space in which alignment must be negotiated.

    Many of these inequities can be understood as forms of informational asymmetry, which is itself a structural expression of power. Shareholders---developers, firms, and regulators---possess privileged access to the design choices, data pipelines, and interpretive resources that determine how an AI system functions. Stakeholders, by contrast, often lack both transparency and recourse: they cannot inspect models, challenge their decisions, or meaningfully influence how objectives are specified. This epistemic imbalance mirrors classical principal-agent asymmetries but extends them into the realm of political economy. In practice, power in AI systems is exercised through control over information---who collects it, who interprets it, and who is rendered visible or invisible within it. Reframing power as informational asymmetry helps integrate structural critiques of inequality directly into the analytical vocabulary of alignment research.

    As suggested above, scaling exacerbates these asymmetries. As models grow in size and complexity, the technical expertise and computational resources required to understand or replicate them concentrate within a small set of institutions. This concentration further widens the informational gap between shareholders and stakeholders, making it more difficult for affected communities to contest design decisions or influence governance. The result is a structural coupling between scale and power: the larger and more widely deployed the system, the more alignment depends on institutional arrangements that determine whose knowledge and whose values shape its behaviour.

    Various forms of human exploitation that influence and shape training and testing environments for AI models designed for real-world deployment can occur either through the direct utilisation of underpaid and poorly trained labour or the collection and utilisation of individuals' data without their consent \citep{Gray-Suri-2019}. In many cases, those most affected by the deployment of AI systems---i.e., the stakeholders---are excluded from decisions about their design, objectives, and governance, which are made primarily by shareholders (developers, corporations, and regulators). Alignment, therefore, requires rebalancing the power dynamics that shape AI development, ensuring that diverse voices have meaningful influence in defining and evaluating alignment objectives \citep{Miceli-et-al-2022}. Seen through the lens of the scaling hypothesis, this requirement is not incidental but structural: as AI systems become more general and widely deployed, alignment increasingly depends on governance mechanisms capable of managing both informational asymmetry and pluralistic disagreement at scale.

\subsection{Benchmarking Degrees of Alignment}

    A benchmark is supposed to measure a model's performance on a task. To create a standard metric for measuring degrees of value alignment---either singular or pluralistic---requires that the standard can be formalised. To benchmark alignment along the objectives axis, one must already formalise the ``true'' objective that the system is meant to approximate. But because objective functions are themselves proxies for those true objectives, such formalisation is rarely possible. In other words, the very mechanism that makes value misalignment possible---the proxy problem---also makes precise measurement of alignment impossible. Moreover, as \citet{Dotan-Milli-2019} argue, benchmarks do not merely measure progress but actively shape what counts as progress. In particular, they highlight that evaluation practices embed normative assumptions about what constitutes a successful model, thereby influencing which research directions gain legitimacy within the field. %
    %%%
    %%%
    %%%
    For simple, well-specified tasks, it may be feasible to evaluate whether a proxy captures the intended goal. However, as tasks become more socially or ethically complex, proxies become less reliable, and the gap between objective functions and true objectives widens. If we had a reliable metric for measuring that gap, we could use it to design a better proxy; but, since we cannot, benchmarking complex alignment problems becomes conceptually circular. Such evaluation regimes can produce self-reinforcing disciplinary dynamics: once a particular model paradigm performs well on widely adopted benchmarks, the benchmark itself incentivises further optimisation for that paradigm, thereby consolidating its dominance. In this way, benchmarks function not only as evaluative tools but also as institutional mechanisms that stabilise certain research trajectories while marginalising alternatives \citep{Dotan-Milli-2019}. The situation worsens when models are continually retrained on data that reflects the effects of their prior deployments, creating feedback loops that distort both the data distribution and the proxy objectives themselves \citep{ONeil-2016}. When these feedback loops interact with entrenched benchmarking practices, they can further entrench specific value-laden assumptions about performance, fairness, or usefulness, making the underlying evaluative framework appear natural or inevitable even though it reflects historically contingent design choices. Under these conditions, misalignment is not only difficult to detect but also self-reinforcing.

    The resulting insight is negative but fundamental: for sufficiently complex objectives, measuring degrees of misalignment is impossible in principle. When this is the case, decisions about whether and how to deploy AI systems must hinge not on technical certainty but on inductive risk---the balance of potential harms and benefits under uncertainty \citep{Hempel-1965, Douglas-2000}. Philosophers of science have argued that scientific inference always involves value-laden judgments about acceptable levels of error and harm. The same holds for AI benchmarking: every metric embodies a decision about what counts as ``good enough'', which errors matter, and to whom. The analysis of value-laden disciplinary shifts makes this point especially salient for ML: evaluation metrics and benchmark suites embed social and political judgments about which capabilities deserve optimisation and which risks are tolerable. Benchmarks, then, are not neutral instruments but normative artefacts that encode trade-offs between competing values---e.g., accuracy versus fairness or efficiency versus accountability. Acknowledging this continuity between epistemology and engineering reframes alignment as a problem of responsible inquiry. The central question shifts from ``Is the system aligned?'' to ``Is the system aligned enough? for whom? and, at what cost?''

\section{Alignment As Ongoing Governance}
    \label{sec:Governance}

    The structural definition of the value alignment problem reveals that alignment is not a singular technical or moral challenge but a class of structurally related problems arising wherever a principal delegates a task to an agent. In human-human contexts, such delegation introduces value misalignment because principals and agents already possess distinct and sometimes competing values. In contrast, artificial agents do not possess intrinsic values: they instantiate proxy objectives designed and optimised by humans. Misalignment, therefore, emerges not from the agent’s will but from the incompleteness, ambiguity, or mis-specification of those proxies.
        %As mentioned, the contrast between human-human and human-AI principal-agent relationships underscores a key insight: in the economic case, values are intrinsic---each agent possesses inherent motivations, goals, and interests. In the AI case, however, objectives are supplied. Hence, one aspect of misalignment arises from the impossibility of perfectly specifying an objective function. The ``values'' of an artificial agent are not natural dispositions but design artifacts.
    %%%
    From this perspective, alignment failures are not anomalies but expectable outcomes of imperfect formalisation under informational and social constraints. The principal-agent analogy underscores that these failures are driven less by error than by structure---by the impossibility of completely encoding human objectives in computational form and by the asymmetries of information that arise between designers, systems, and affected communities.
        %From this perspective, the structural value alignment problem  reframes the central question of alignment research: whose values are we aligning to? This question moves the value alignment problem beyond its purely technical or normative dimensions into a fundamentally {\it social} domain. Alignment is not simply about modelling human values correctly---it is about negotiating among conflicting human interests within power-laden institutional structures.

    Pluralistic alignment reframes the central question: whose values are we aligning to, and who decides? Once we recognise that there are many principals---shareholders and stakeholders with diverse, and often conflicting, objectives---alignment ceases to be a matter of fitting models to ``the values of humanity''. It becomes, instead, a question of institutional design and power. The structural framework exposes this dynamic: even perfect proxy objectives cannot guarantee global alignment when values diverge across communities or social strata. A system that is well-aligned with the goals of its shareholders may be deeply misaligned with the interests of its stakeholders. Thus, the alignment problem is not only technical or normative, but fundamentally social and political in nature. %
        %The principal-agent analogy also helps clarify why misalignment persists. The problem arises not merely from value differences but from informational asymmetries: the principal (human) cannot fully observe, understand, or control the agent's (AI's) internal processes. Moreover, objective functions---proxies for our true goals---can only ever imperfectly encode human intentions. From the system's perspective, these proxies are the goals themselves. The resulting gap between our objectives and the system’s operational definition of those objectives is the fundamental source of the value alignment problem.
    %%%
    %%%
    %%%
    On the structural view, alignment must be understood as a problem of governance. Because misalignment can arise independently along the objectives, information, and principals axes, any attempt to mitigate it must be implemented through institutions capable of regulating each axis in practice. The three-axis framework serves not only as an analytic taxonomy but as a guide to institutional design: it identifies distinct failure modes, clarifies which actors must be empowered to address them, and explains why interventions that solve one aspect of alignment may worsen misalignment along another.

    In practice, alignment depends not on any single mechanism but on the interaction of these institutional features across all three axes. Institutions addressing the objectives axis determine which goals AI systems are permitted to optimise and under what constraints. These may include internal safety review boards, external regulatory agencies, standards-setting bodies, and liability regimes that are able to shape the incentives of developers and deployers. Institutions addressing the information axis aim to reduce epistemic asymmetries between those who build systems, those who govern them, and those affected by them. Transparency requirements, auditing mandates, interpretability standards, incident reporting obligations, and independent evaluation organisations all function by redistributing information that would otherwise remain concentrated in the hands of system designers. Institutions addressing the principals axis determine whose interests are treated as authoritative when trade-offs must be made. Public consultation procedures, stakeholder representation, collective feedback mechanisms, democratic oversight, and participatory design processes all operate by expanding or redefining the set of principals whose values are taken to matter. A system's behaviour depends not only on its training objective but on who chose that objective, who was excluded from the decision, who can inspect the resulting model, and who has the authority to contest its deployment. The three-axis framework makes these dependencies explicit and allows alignment research to reason systematically about the design and implementation of governance structures rather than treating institutional context as background noise.
    
    A practical implication of this view is a shift from the rhetoric of ``solving'' value alignment to one of managing and mitigating misalignment across its three axes. Misalignment scales with model size, data scope, and computational power---the very factors that define modern deep-learning approaches to AI. As models become more general-purpose, informational asymmetries deepen, and pluralistic divergence widens. Perfect alignment, even in principle, becomes impossible. What remains is the task of designing procedures and institutions capable of continuously negotiating, auditing, and correcting misalignment as it arises.
        %However, the structural analogy to the principal-agent framework also introduces a potential advantage: unlike human agents, artificial agents can, in principle, be designed from the ground up. We can shape their objectives, modify their incentives, and encode their learning environments. This design flexibility means that the central challenge of alignment is not merely technical but institutional---how to ensure that the processes by which objectives are defined and encoded reflect the legitimate interests of all relevant principals.

    This structural account also clarifies the limitations of understanding alignment as consisting of two interrelated components: a technical problem (how to encode values) and a normative problem (which values to encode). While this distinction captures important conceptual differences, it implicitly assumes that once the correct values are identified and correctly specified, alignment will follow. (Suggesting, also, that it is {\it possible} to identify the ``correct'' values for AI.) The three-axis framework shows that this assumption is false. Both the technical and the normative components of value alignment correspond primarily to the objectives axis, with the normative problem determining the target objective and the technical problem determining how to implement it. However, failures along the information axis and the principals axis cannot be reduced to either technical difficulty or moral disagreement. Instead,they arise from organisational incentives, resource asymmetries, and institutional secrecy. Opacity, unverifiability, and asymmetric access to data are not simply technical limitations but features of organisational arrangements, incentives, and resource distribution. Likewise, failures along the principals axis are not merely disagreements about values in the abstract but conflicts over authority, representation, and decision-making power. The technical/normative distinction therefore collapses multiple structurally distinct problems into a single dimension, whereas the three-axis decomposition embeds this distinction within a broader socio-technical analysis that makes explicit the institutional conditions under which alignment succeeds or fails.
    %%%
    %%%
    This reconceptualisation, therefore, situates alignment within a broader frame of socio-technical governance. AI systems are embedded in networks of labour, data extraction, and environmental cost. These systems' ``objectives'' are operationalised not only through code but through economic incentives, regulatory regimes, and the social hierarchies in which they are deployed. To treat alignment as an engineering problem alone is to ignore the upstream power structures that determine who gets to specify objectives and who bears the risks when they fail. %
        %Alignment, on this structural description, therefore compels a broader re-framing of the research agenda: from the pursuit of a universal ``solution'' to the development of ongoing mechanisms of negotiation and governance. AI alignment, on this view, is not an engineering problem to be solved once and for all, but a continuous process of managing information, objectives, and power across evolving socio-technical systems.
    %%%

    Recent work on ``fair process'' views of alignment suggest that alignment may sometimes be achieved through agreement on {\it procedures} rather than through agreement on values \citep{Gabriel-Keeling-2025}. On this view, alignment does not require agreement on a single set of values; instead, it requires agreement on a decision procedure that stakeholders regard as legitimate. If the process by which objectives are defined is inclusive, transparent, and procedurally fair, then the resulting system may count as aligned even when substantive disagreement persists. Hence, pluralism is managed by institutionalising fair procedures rather than by identifying a universally correct objective. This suggests a possible alternative to the structural view: alignment might be achieved by securing consensus at the level of process, allowing the principals axis to be stabilised through procedural legitimacy. The structural framework, however, predicts that fair-process solutions {\it alone} cannot fully resolve the alignment problem, because agreement on procedure does not eliminate the other sources of misalignment. Even when stakeholders agree on a decision procedure, objective proxies may still be imperfect, and informational asymmetries may still prevent meaningful oversight. Moreover, it is unclear whether complete procedural legitimacy is possible in real-world institutions because differences in power, expertise, and access to information mean that some actors inevitably exercise greater influence over the design and deployment of systems. For this reason, fair-process accounts should be understood not as replacements for the structural analysis but as governance mechanisms operating primarily along the principals axis. They can reduce conflict over authority without eliminating misspecification or opacity, and the three-axis framework makes these limits visible

    Consequently, alignment demands institutional mechanisms for oversight, participatory design, and con\-tes\-tab\-il\-ity---mechanisms that can adapt as both models and social conditions evolve. The challenge is not to achieve perfect convergence between human and machine values, but to build systems resilient to divergence: systems capable of being corrected, contested, and re-aligned as our understanding of ``our values'' itself changes. Under this interpretation, governance is not an external constraint on alignment but its primary medium. The task of alignment research therefore expands from designing better objective functions to designing better decision procedures, oversight structures, and forms of collective control. The three-axis decomposition provides a framework for this task by allowing us to ask, for any proposed system: which objectives does it optimise, who can understand and evaluate it, and whose interests determine whether it should exist at all?

    In this light, the question for future alignment research is not merely how to make AI systems safe or compliant, but how to make them responsive to human needs. If the objective of alignment is to ensure that AI systems act in accordance with human values, then the primary task becomes identifying which humans, which values, and through what processes those values are continually negotiated and maintained. Alignment, properly understood, is not an endpoint of technical control---it is the architecture of shared agency. Considering the objectives encoded in AI systems with respect to a particular set of principals sheds light on how AI systems fail to satisfy these objectives. In general, when any given AI model is touted as a solution---particularly by the shareholders of that system---it is fruitful to ask: to what problem?

%\begin{acks}
%    \ldots
%\end{acks}

\section*{Generative AI Usage Statement}

The author(s) did not use generative AI in the writing of this paper.

\bibliographystyle{ACM-Reference-Format}
\bibliography{Biblio}

%\appendix



\end{document}